# SPECTRAL ANALYSIS OF NEUTRINO MIXING MECHANISMS

C. N. LEUNG[1]

*Department of Physics and Astronomy, University of Delaware,*
*Newark, DE 19716*

and

P. I. KRASTEV

*Institute of Field Physics, Department of Physics and Astronomy,*
*The University of North Carolina at Chapel Hill, CB-3255, Phillips Hall,*
*Chapel Hill, NC 27599-3255*

## ABSTRACT

The prospects of using spectral measurements of solar neutrinos to distinguish different mechanisms of neutrino flavor mixing are analyzed. The mechanisms studied include the conventional mass mixing mechanism and an alternative mechanism which assumes a flavor nondiagonal neutrino-gravity coupling.

A mechanism of neutrino flavor mixing was proposed[1,2] several years ago which assumes that neutrinos couple to gravity in a flavor nondiagonal manner. Such a coupling constitutes a violation of the equivalence principle (EP). Consequently, experiments such as the solar neutrino experiments which measure neutrino flavor mixing can be used to test whether neutrinos obey the EP. It was shown[3] that existing solar neutrino data probed the EP at the level of few parts in $10^{14}$, which is better than the best limit[4] (for ordinary matter) obtained from torsion balance experiments by more than an order of magnitude.

In this mechanism, which shall be referred to as the G-mechanism, the neutrino weak interaction (or flavor) eigenstates are assumed to be distinct from their gravitational interaction eigenstates and that they can be expressed as linear superpositions of the gravitational eigenstates, with a mixing angle $\theta_G$ (we assume two neutrino flavors for simplicity). It is further assumed that the two gravitational eigenstates couple to gravity with different strength, thus violating the EP and leading to neutrino flavor oscillations when a neutrino propagates in a gravitational field. We stress that this mechanism does not require neutrinos to have a mass. The evolution equations for relativistic flavor neutrinos propagating in a weak gravitational field are given by

$$i\frac{d}{dr}\begin{pmatrix}\nu_e\\\nu_\mu\end{pmatrix} = -2E_\nu\phi(r)\Delta f\begin{bmatrix}0 & \frac{1}{2}\sin 2\theta_G\\\frac{1}{2}\sin 2\theta_G & \cos 2\theta_G\end{bmatrix}\begin{pmatrix}\nu_e\\\nu_\mu\end{pmatrix} \qquad (1)$$

---

[1]Presented by C. N. Leung at the DPF'94 Meeting in Albuquerque, New Mexico, August 2 - 6, 1994.

where $E_\nu$ is the neutrino energy, $\phi(r)$ is the Newtonian gravitational potential (we assume a static, spherically symmetric source for the gravitational field), and $\Delta f \equiv f_2 - f_1$ is a measure of the degree of EP violation. The parameters $f_{1,2}$ can be identified as parameters in the Parametrized Post-Newtonian formalism.[5] In general relativity, $f_1 = f_2 = 1$.

Eqs. 1 have the same form as the evolution equations for neutrinos propagating in vacuum in the familiar mass mixing mechanism of Pontecorvo[6] (M-mechanism for short) except that in the M-mechanism,

$$\theta_G \to \theta \quad \text{and} \quad -2E_\nu \phi(r) \Delta f \to \frac{\Delta m^2}{2E_\nu}, \tag{2}$$

where $\theta$ is the vacuum mixing angle relating the neutrino flavor eigenstates to their mass eigenstates and $\Delta m^2 \equiv m_2^2 - m_1^2$ is the neutrino mass-squared difference. As a result of Eq. 2, the two mechanisms lead to a different energy dependence in the oscillating term of the neutrino survival probability. In the M-mechanism, the neutrino oscillation length is proportional to the neutrino energy: $\lambda_M = 4\pi E_\nu / \Delta m^2$, whereas in the G-mechanism the oscillation length (for a constant gravitational potential) is inversely proportional to $E_\nu$: $\lambda_G = \pi/(E_\nu |\phi| \Delta f)$.

Unlike the M-mechanism, the possibility of explaining the solar neutrino data by long-wavelength vacuum oscillations in the G-mechanism is ruled out at the $3\sigma$ level.[3] On the other hand, if we incorporate the MSW mechanism[7] of resonant transitions in the Sun, our $\chi^2$-analysis of the most recent solar neutrino data reveal two allowed regions in the $\Delta f - \sin^2(2\theta_G)$ plane[8]: a "small mixing region" for $2 \times 10^{-3} \leq \sin^2(2\theta_G) \leq 10^{-2}$ and $2.7 \times 10^{-14} \leq \Delta f \leq 3.1 \times 10^{-14}$; and a "large mixing region" for $0.6 \leq \sin^2(2\theta_G) \leq 0.9$ and $1.0 \times 10^{-16} \leq \Delta f \leq 1.5 \times 10^{-15}$, at 95% C.L. These allowed regions are compatible with the ones found earlier in Ref. 3 and disagree with the findings of Ref. 9.

The evolution equations for neutrinos propagating in the Sun are obtained from Eqs. 1 by replacing $\cos 2\theta_{(G)}$ with $\cos 2\theta_{(G)} - \frac{\sqrt{2}G_F N_e(r)}{2\pi/\lambda_{M(G)}}$, for the M (G)-mechanism. Here $G_F$ is the Fermi constant and $N_e(r)$ is the electron number density inside the Sun. For the G-mechanism, a resonance occurs when

$$E_\nu = \frac{\sqrt{2}G_F N_e(r)}{2|\phi(r)|\Delta f \cos 2\theta_G} \tag{3}$$

and the transition is adiabatic if

$$\frac{\sqrt{2}G_F (N_e)_{res} \tan^2(2\theta_G)}{\left|\left(\frac{1}{N_e}\frac{dN_e}{dr}\right) - \left(\frac{1}{\phi}\frac{d\phi}{dr}\right)\right|_{res}} \gg 1. \tag{4}$$

The adiabaticity condition for the M-mechanism can be obtained from Eq. 4 by neglecting the $\frac{1}{\phi}\frac{d\phi}{dr}$ term in the denominator and replacing $\theta_G$ by $\theta$. Note that, in contrast to the M-mechanism, adiabatic transitions in the G-mechanism requires high energy neutrinos, e.g., $E_\nu \Delta f >$ few $\times 10^{-13}$ MeV, for small mixing angles. It is this different energy dependence which leads to distinguishable spectral predictions for the two mechanisms.

We display in Fig. 1a the $^8$B solar neutrino spectrum ($F(E_\nu)$) predicted for the G-mechanism. The spectrum is normalized to the standard (no mixing) $^8$B-neutrino spectrum ($F_{st}(E_\nu)$) such that this ratio is 1 at $E_\nu = 10$ MeV. For comparison we show also the corresponding spectral distortions in the M-mechanism for both MSW transitions (Fig. 1b) and vacuum oscillations (Fig. 1c). For large mixing angles the G-mechanism yields little distortion of the spectrum, which makes it impossible to distinguish it from the MSW case of the M-mechanism. On the other hand, for small mixing the G-mechanism leads to a much more substantial distortion at the high energy end of the spectrum than the M-mechanism. This will allow for future solar neutrino experiments such as SNO to differentiate it from the M-mechanism, as indicated by the estimated error bars after 5 years of operation of the SNO detector (Efficiency and energy resolution have not been included in the estimate of these error bars). A more detailed discussion of these spectra can be found in Ref. 8.

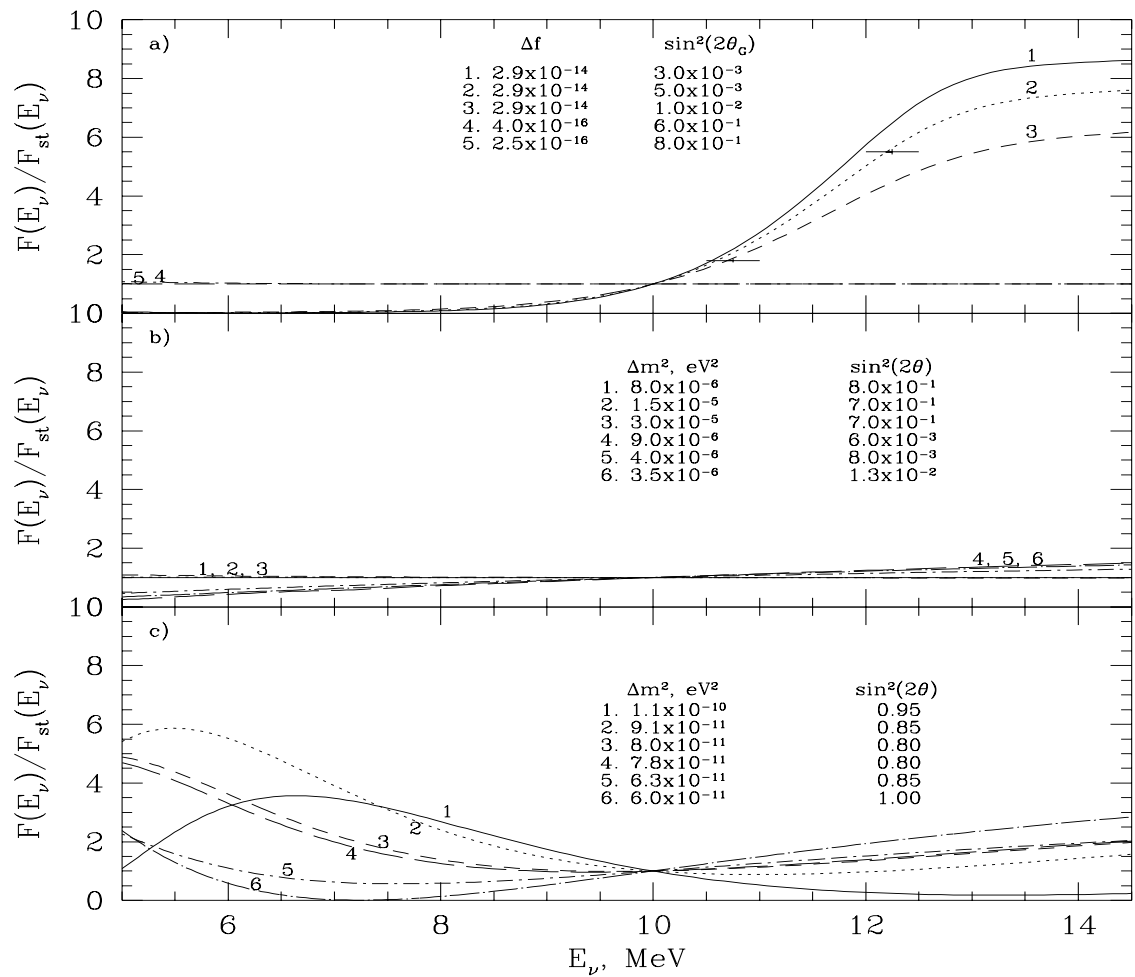

Fig. 1. $^8$B–neutrino spectra for: a) G-mechanism, b) MSW effect in the M-mechanism, and c) vacuum oscillations in the M-mechanism. The chosen values of the parameters for each case correspond to values allowed by the current solar neutrino data.

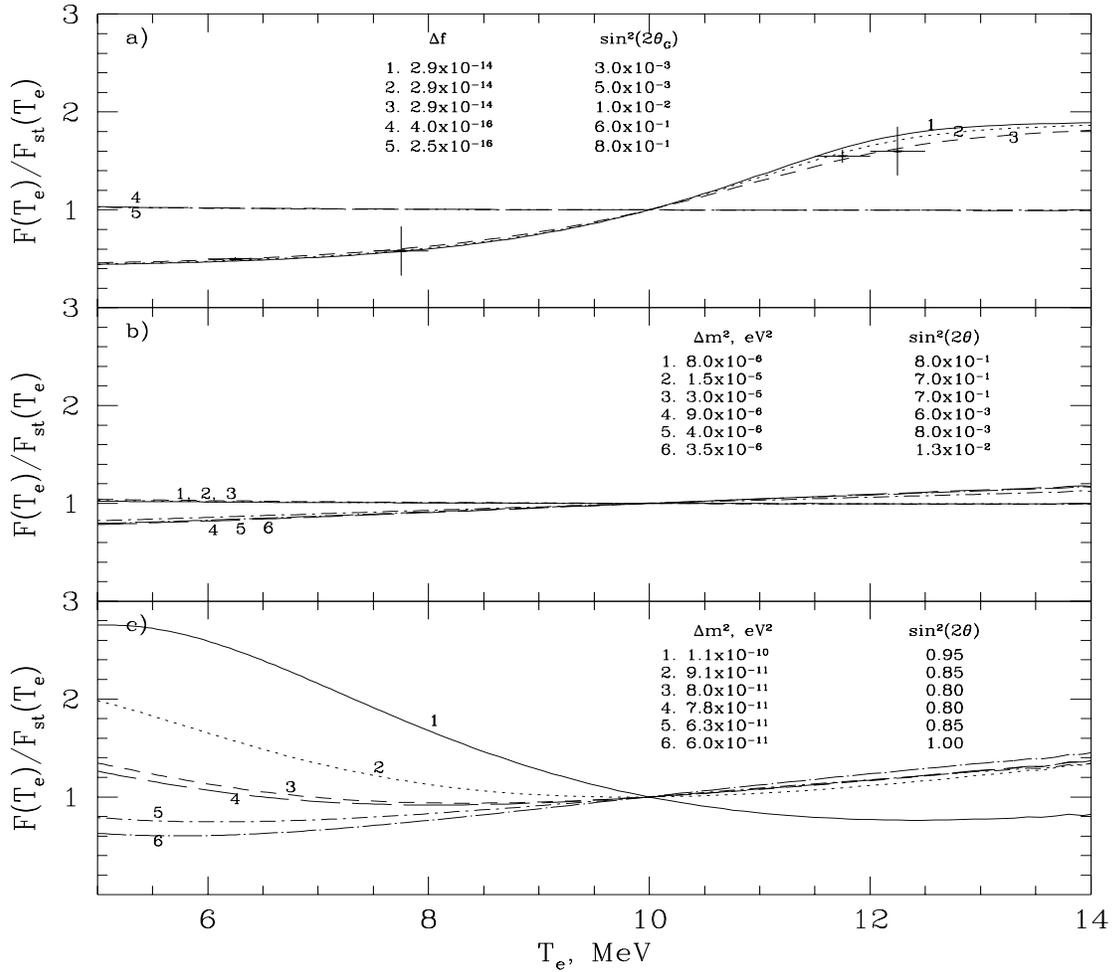

Fig. 2. Recoil–electron spectra for the same three cases as in Fig. 1.

We have also studied the prospects of using the spectra of recoiled electrons in $\nu e$-scattering (e.g., in SuperKamiokande) to distinguish the G-mechanism from the M-mechanism. Our results displayed in Fig. 2 show that the spectral distortions are much less prominant, which makes it more difficult to distinguish the two mechanisms in this case.

In summary, the G-mechanism is currently a viable solution to the solar neutrino problem. If the mixing angles are small, a careful measurement of the solar neutrino spectrum in the upcoming experiments will be able to determine which is the underlying mechanism responsible for the observed solar neutrino deficit.